\providecommand{\jocg}{0}
\ifnum \jocg=0
  \documentclass[11pt]{article}
  \newcommand{\titleformat}[1]{#1}
\else
  \documentclass[11pt]{jocg}
  \newcommand{\titleformat}[1]{\MakeUppercase{#1}}
\fi

\usepackage{amsfonts,amssymb,amsthm,eucal,amsmath}
\usepackage{graphicx}
\usepackage[T1]{fontenc}
\usepackage{latexsym,url}
\usepackage{array}
\usepackage{subfig}
\usepackage{comment}
\usepackage{color}
\usepackage{hyperref}
\usepackage{cprotect}
\usepackage{lineno}

\usepackage[inner=30mm, outer=30mm, textheight=225mm]{geometry}

\newcommand{\R}{\mathbb{R}}

\newcommand{\rand}{\operatorname{rand}}

\newcommand{\halpha}{\hat{\alpha}}
\newcommand{\hbeta}{\hat{\beta}}

\title{\titleformat{A deterministic pseudorandom perturbation scheme for arbitrary polynomial predicates}}
\ifnum \jocg=0
  \author{Geoffrey Irving\thanks{Email: \{irving,forrest\}@otherlab.com, Otherlab, San Francisco, CA, United States}
  \and Forrest Green$^*$}
  \date{Version 1, \today}
\else
  \author{Geoffrey Irving,%
          \thanks{\affil{Otherlab},
                  \email{\{irving,forrest\}@otherlab.com}}\,
          Forrest Green\footnotemark[1]}
\fi

\begin{document}
\ifnum \jocg=1
  \linenumbers
\fi
\maketitle

\begin{abstract}
We present a symbolic perturbation scheme for arbitrary polynomial geometric predicates which combines the benefits of
Emiris and Canny's simple randomized linear perturbation scheme with Yap's multiple infinitesimal scheme for general predicates.
Like the randomized scheme, our method accepts black box polynomial functions as input.  For nonmaliciously chosen predicates,
our method is as fast as the linear scheme, scaling reasonably with the degree of the polynomial even for fully
degenerate input.  Like Yap's scheme, the computed sign is deterministic, never requiring an algorithmic restart (assuming a
high quality pseudorandom generator), and works for arbitrary predicates with no knowledge of their structure.  We also apply
our technique to exactly or nearly exactly rounded constructions that work correctly for degenerate input, using l'H\^opital's
rule to compute the necessary singular limits.  We provide an open source prototype implementation including example
algorithms for Delaunay triangulation and Boolean operations on polygons and circular arcs in the plane.
\end{abstract}

\section{Introduction}

Symbolic perturbation is a standard technique in computational geometry for avoiding degeneracies by
adding an infinitesimally small perturbation to the inputs of a geometric algorithm.  The technique was introduced by
\cite{edelsbrunner1990simulation}, with refinements in \cite{yap1990symbolic}, \cite{emiris1992efficient}, \cite{emiris1995general},
and \cite{seidel1998nature}.  Consider a geometric function $G : \R^N \to S$ mapping input coordinates $x \in \R^N$ into
some discrete set $S$.  Examples of $G(x)$ include Delaunay triangulation, arrangements of lines or circles, and Boolean operations
on shapes.  We will assume $G(x)$ can be computed using an algorithm that queries its input $x$ only through the
signs of various polynomials $f(x)$ with integer coefficients, each representing a geometric predicate such as
``is this triangle counterclockwise?'' or ``do two circles intersect inside a third circle?''.  If $f(x) = 0$, the algorithm
either fails due to ambiguity or requires special logic to handle the degeneracy.

We describe symbolic perturbation in the framework of nonstandard analysis; see \cite{yap1990symbolic}, \cite{emiris1995general},
and \cite{seidel1998nature} for the geometric meaning of this approach.  To extend $G(x)$ to degenerate inputs, we introduce one
or more positive infinitesimal quantities $\epsilon_1, \epsilon_2, \ldots$, with $0 < \epsilon_i < 1/n$ for all $i,n > 0$.  If
we introduce more than one infinitesimal, we define a relative ordering of the different monomials $\epsilon_1^{p_1} \epsilon_2^{p_2} \cdots$;
the simplest is lexicographic ordering where $\epsilon_i^p > \epsilon_{i+1}$ for all $i,p > 0$.
We then form an infinitesimal perturbation $\delta \in \R[\epsilon_1,\epsilon_2, \ldots]^N$ from
linear combinations of the infinitesimals (here $\R[\epsilon_i]$ is the ring of multivariate polynomials over $\R$ generated by $\epsilon_i$), and evaluate
\begin{linenomath*}
\begin{align*}
G'(x) = G(x+\delta).
\end{align*}
\end{linenomath*}
In detail, whenever the algorithm asks for the sign of $f(x)$ for some integer coefficient polynomial $f$, we instead compute
$f(x+\delta)$, which is a multivariate polynomial in the infinitesimals.  The sign of $f(x+\delta)$ is the sign of the ``least infinitesimal''
nonzero monomial coefficient of this polynomial.
We distinguish between three existing symbolic perturbation schemes that can be expressed in this framework and discuss their advantages
and disadvantages.

{\bf Yap's deterministic scheme} \cite{yap1990symbolic} introduces one infinitesimal $\epsilon_i$ per input coordinate $x_i$, and lets $\delta_i = \epsilon_i$.  This
corresponds to evaluating $f(x_1+\epsilon_1, x_2+\epsilon_2, \ldots)$.  Since each coordinate has its own infinitesimal, $f(x+\delta)$ has at least
one nonzero monomial unless $f$ is identically zero, so the scheme produces a nonzero sign for all nonzero polynomials.  Unfortunately, a degree $d$ polynomial
$f$ results in an $f(x+\delta) \in \R[\epsilon_1,\epsilon_2, \ldots]$ with up to $\binom{n + d}{n}$ monomial terms where $n$ is the number of input
coordinates used by $f$, which is worst case exponential in the degree of the predicate.  For extremely degenerate input, we may need to evaluate a large
number of coefficients before finding a nonzero.

{\bf Emiris and Canny's deterministic linear scheme} \cite{emiris1992efficient} arranges the input coordinates into $n$ $k$-vectors based on the dimension
$k$ of the geometric space as $x_{a,b}$, $1 \le a \le n$, $1 \le b \le k$.  They introduce a single infinitesimal $\epsilon$ and write
\begin{linenomath*}
$$\delta_{a,b} = \epsilon \cdot (a^b \operatorname{mod} p)$$
\end{linenomath*}
where $p > n$ is a prime.  They show that this scheme produces a nonzero sign for simplex orientation tests up to dimension $k$ and for
the incircle tests used in Delaunay triangulation.  However, as discussed in \cite{seidel1998nature}, extending this technique to other predicates
is difficult.

In addition, as noted in \cite{burnikel1994degeneracy}, a fixed deterministic perturbation may turn highly degenerate input into worst case behavior for algorithms
like convex hull: ignoring the $\operatorname{mod} p$, the deterministic linear scheme produces a convex hull of size $n^{\lceil d/2 \rceil}$ when all input
points are at the origin.  We believe this also applies to Yap's scheme and may arise with the modular deterministic linear scheme.

{\bf Emiris and Canny's randomized linear scheme} \cite{emiris1995general} again introduces a single infinitesimal $\epsilon$, but now sets
$\delta_i = \epsilon y_i$ using random coefficients $y_i$ chosen from some space $Y$.  By the Schwartz-Zippel lemma \cite{schwartz1980fast}, $f(x+\delta)$
will be nonvanishing as a polynomial in $\epsilon$ with probability at least $1 - d/|Y|$, where $d$ is the degree of the polynomial.  Unfortunately, what we
actually need is for \emph{all} polynomials evaluated during the algorithm to not vanish, which reduces the probability of success to
$(1 - d/|Y|)^T$ where $T$ is the number of branches required.  Emiris and Canny show that their randomized scheme is very efficient in the
algebraic computation model, but suffers from a worst case cubic slowdown in the bit computation model due to the large $|Y|$ required.  For some algorithms
it is possible to reduce this slowdown by restarting only part of the algorithm, but this adds significant complexity (in the authors' experience).

To summarize: Yap's deterministic scheme and the randomized linear scheme work for arbitrary polynomial predicates, but suffer from unfortunate performance penalties.
The randomized linear scheme occasionally requires a restart of all or part of the computation, adding extra complexity to the surrounding algorithm
especially if multiple computations are chained together (possibly with user interaction in between).  The deterministic linear scheme is ideal when it works
but requires special analysis to verify correctness for each predicate.

Our contribution is to combine the advantages of each of the above methods.

\section{A deterministic pseudorandom perturbation}

Our approach is to introduce an infinite sequence of infinitesimals $\epsilon_1, \epsilon_2, \ldots$, choose deterministic pseudorandom vectors $y_1, y_2, \ldots$
with $y_{k,i} = \rand(k,i)$ for $1 \le k < \infty, 1 \le i \le n$, and set
\begin{linenomath*}
$$\delta = \epsilon_1 y_1 + \epsilon_2 y_2 + \cdots.$$
\end{linenomath*}
Here $\rand$ is a deterministic pseudorandom generator with random access capability.  Our implementation uses the Threefry generator of
\cite{salmon2011random}, with
\begin{linenomath*}
$$\rand : [0,2^{128}) \times [0,2^{128}) \to [0,2^{32}).$$
\end{linenomath*}
We order the infinitesimals largest first, so that $\epsilon_i^p > \epsilon_{i+1}$ for all $p > 0$.  As in Yap's scheme, this ordering lets us add one term of the
perturbation series at a time, evaluating
\begin{linenomath*}
\begin{align*}
f_0 &= f(x) \\
f_1 &= f(x + \epsilon_1 y_1) \\
f_2 &= f(x + \epsilon_1 y_1 + \epsilon_2 y_2) \\
f_3 &= f(x + \epsilon_1 y_1 + \epsilon_2 y_2 + \epsilon_3 y_3) \\
\vdots
\end{align*}
\end{linenomath*}
and stopping as soon as we arrive at a nonzero polynomial $f_k(\epsilon_1. \ldots, \epsilon_k)$.
To compute the coefficients of a given $f_k$, we temporarily view the infinitesimals $\epsilon_i$ as integer variables and use a black box function for $f(x)$ to evaluate
$f_k(\epsilon_1, \ldots, \epsilon_k)$ with $(\epsilon_1, \ldots, \epsilon_k)$ replaced with all $\binom{k+d}{k}$ nonnegative integer tuples satisfying $\epsilon_1 + \cdots + \epsilon_k \le d$
as discussed in \cite{neidinger2009multivariable} and \autoref{polynomial}.
If any values are nonzero, we use multivariate polynomial interpolation to recover the $\binom{k+d}{k}$ coefficients of $f_k$ and return the sign of the least infinitesimal nonzero term.
Note that we have replaced the $\binom{n+d}{n}$ coefficients of Yap's scheme with $\binom{k+d}{k}$ coefficients.

We show that the computational cost is dominated by the first perturbation term even for arbitrarily degenerate input, as long as the range $Y$ of the random generator
satisfies $d^3 \ll |Y|$.  In other words, our scheme has the same cost as the simple linear scheme.  To see this, note that if $f_k$ is zero, setting one $\epsilon_j$ to one and the others to zero
shows that $f(x + y_1), \ldots, f(x + y_k)$ are zero.
Thus, if the polynomial predicate $f(x)$ is not identically zero, the Schwartz-Zippel lemma gives
\begin{linenomath*}
$$\Pr(f_k = 0) \le \frac{d^k}{|Y|^k}.$$
\end{linenomath*}
The sizes of the lattice points on which we evaluate $f$ grow slowly with $k$, so the cost of a single polynomial evaluation is effectively $O(1)$ where the constant depends on the
polynomial.  Similarly, the sizes of the numbers used for multivariate interpolation also grow slowly with $k$, so the cost of multivariate
interpolation at level $k$ is $O\left(d \binom{d + k}{k}^2\right)$ (see \autoref{polynomial}).  Thus, the expected cost of the perturbation scheme is
\begin{linenomath*}
\begin{align*}
\sum_{k = 0}^\infty \Pr(f_k = 0) O\hspace{-.25em}\left(d \binom{d + k + 1}{k + 1}^2\right) \le \sum_{k = 0}^\infty \frac{d^k}{|Y|^k} O(d^{2k+3})
  = O\hspace{-.25em}\left(d^3 \sum_{k = 0}^\infty \frac{d^{3k}}{|Y|^k} \right) = O(d^3)
\end{align*}
\end{linenomath*}
where we need $d^3 < |Y|$ to guarantee a convergent geometric series.  In practice, $d^3 \ll |Y|$; for $|Y| = 2^{32}$ terms with $k \ge 2$ contribute less than $1/4000$th of the expected cost
for polynomials up to degree $100$.  We emphasize that this bound is independent of the input $x$, and therefore holds even for maliciously chosen input data.  However, we do
assume that $\rand$ behaves as a strong random source and, in particular, that the polynomials $f(x)$ are not chosen with knowledge of $\rand$.\footnote{Though maliciously
choosing $f(x)$ so that $f_1 = f_2 = 0$ is quite useful for unit testing purposes.}

Thus, our method has the same complexity as the deterministic linear scheme, but like Yap's scheme and the randomized linear scheme it works on arbitrary polynomials.  As in the
randomized scheme, the perturbation does not create any worst case behavior not already present in the input data.  Since the occasional random fallbacks occur one
polynomial at a time, the outer structure of a geometric algorithm is blissfully unaware that randomness is used internally, and in particular we avoid poor bit complexity
scaling when evaluating many predicates over the course of an algorithm.

In practice, the dominant cost of the algorithm is black box predicate evaluation.  Even a single multiplication of two degree $d/2$ terms has complexity $O(d^2)$ using naive
quadratic multiplication (which is typically the fastest algorithm for small degrees).  The linear perturbation phase performs $d$ polynomial evaluations, for a total
complexity of $O(d^3)$, and the constant is typically higher than for interpolation since most polynomials involve several such multiplications.  An $O(d^3)$ slowdown for degenerate
cases is faster than previous general approaches but still a significant drawback (see \autoref{sec:results} for benchmarks).  Fortunately, a tiny amount of finite perturbation
applied to the input can minimize both the $O(d^3)$ slowdown of perturbation and the $O(d^2)$ slowdown of unperturbed exact evaluation, relying on symbolic perturbation to
unconditionally correctly handle the few remaining degeneracies.

\section{Other approaches}

Since the original introduction of the symbolic perturbation method several alternative schemes have been introduced for treating degeneracies in numerical algorithms.
All of these approaches seem to require some algorithm or predicate specific treatment, which complicates the process of developing and especially testing new algorithms.
However, the algorithm specific approaches may be superior to a general approach such as ours when they apply, either by avoiding the slowdown of occasional exact arithmetic
entirely by treating degenerate cases faster (our approach introduces a slowdown of $O(d)$ for the first perturbation level over exact evaluation), or by computing the true exact
answer rather than a perturbed answer.

Perhaps the most natural approach to treating degeneracies is to manually extend the definition of $G(x)$ to degenerate cases and write algorithms which treat these cases
directly.  For example, in an arrangement of lines, intersections of three or more lines can be detected and represented as higher degree vertices in the arrangement graph.
Burnikel et al.\ \cite{burnikel1994degeneracy} argue that perturbation is slower and more complicated to implement than simply handling degeneracies directly and present two degeneracy-aware
algorithms as evidence.  We believe our method reduces the implementation complexity of symbolic perturbation, but agree that a tailored algorithm is faster on highly
degenerate input.  Unlike the deterministic symbolic perturbation schemes, an algorithm built on our method will treat fully degenerate data as purely random data, in particular
avoiding the worst case behavior of convex hull discussed in \cite{burnikel1994degeneracy}.

The \emph{controlled perturbation} approach of \cite{halperin1998perturbation} applies a small finite perturbation to the input points to avoid degeneracies, allowing the rest
of the algorithm to run with inexact floating point arithmetic.  Input points (spheres in their case) are processed one at a time, perturbing each new input to avoid degeneracies
against all previous inputs.  Controlled perturbation requires a careful enumeration of the possible degeneracies that may arise, and a careful choice of the finite tolerance
required for the algorithm to run safely.  A good tolerance bound may be computed with numerical analysis techniques as in \cite{halperin2004controlled}, at the cost of significant
algorithm-specific analysis.  The main advantage of their approach over ours is speed: the majority of their algorithms avoid all exact arithmetic and even all interval arithmetic
or other filters.  As noted above, if degeneracies are pervasive and a slowdown of $O(d^3)$ is too large, an input to a symbolically perturbed algorithm can be randomly
jittered by a small amount, reducing the practical overhead to the cost of interval analysis filtering without affecting correctness.  Unlike controlled perturbation, this requires
no algorithm specific analysis. 

Devillers et al.\ \cite{devillers2012qualitative} present \emph{qualitative symbolic perturbation}, which replaces the algebraic perturbations used in previous perturbation schemes (and ours)
by a sequence of carefully chosen, geometrically meaningful perturbations.  Their approach replaces the $O(d)$ slowdown of the first perturbation level with a predicate dependent
slowdown and may be faster than our method when it applies.  However, the geometric perturbations and the analysis of their effect on the predicates must be performed separately for each
predicate, which complicates the design of algorithms and is a likely source of complexity during implementation and debugging.  Moreover, since the perturbations depend on the
algorithm, chaining two algorithms together requires adjusting the perturbations to be compatible.  Their approach shares with ours (and indeed
with Yap's) the idea of a sequence of increasingly small perturbations, applied one at a time until a nonsingular result is obtained.

Finally, we address a common complaint against symbolic perturbation (e.g., \cite{burnikel1994degeneracy}), namely that a complicated postprocessing step is required to obtain the
exact answer from the perturbed result.  We argue that the input to a typical geometric algorithm already contains some degree of noise or numerical inaccuracy, and therefore that
classes of errors arising from infinitesimal symbolic perturbation already arise in practice for exact algorithms run on slightly bad input data.  For example, consider the Boolean
union of two squares which touch exactly along one edge.  An exact algorithm run on this ideal input would merge the two squares into one rectangle, while symbolic perturbation may
leave the squares separate or even join them only partway along the edge.  However, if the input is already slightly shifted, both algorithms produce exactly the same result.
The solution in both cases is to offset the squares slightly outwards prior to union, which resolves both infinitesimal and finite errors.

\section{Implementation}

A C++ implementation of our symbolic perturbation technique is available under a BSD license at \url{https://github.com/otherlab/core/tree/exact}\footnote{See 
\url{https://github.com/otherlab/core/commit/dc0f10918d17507d} for the version benchmarked below.}.  The code includes three algorithms built on top of the perturbation
core: Delaunay triangulation, Boolean operations on polygons, and Boolean operations on polygons built from circular arcs.  We plan to expand the set of implemented algorithms
and use them for various tasks in CAD/CAM such as shape decomposition for manufacturing and motion planning.  Benchmarks and plotting scripts are available along with the
paper source at \url{https://github.com/otherlab/perturb}.

For simplicity and speed, our implementation quantizes all input coordinates to the integer range $[-2^{53},2^{53}]$, the largest range of integers exactly representable in
double precision.  This allows use of fast interval arithmetic filters \cite{bronnimann2001interval}, falling back to exact integer evaluation using GMP if the filter fails \cite{granlund2012gmp},
and falling back to symbolic perturbation if the exact answer is zero.  The polynomial is provided as a black box evaluation routine (see \verb+exact/perturb.h+ in the code).  For multivariate
interpolation we evaluate $f_k(\epsilon_1, \ldots, \epsilon_k)$ on our fixed set of $(\epsilon_1, \ldots, \epsilon_k)$ tuples, use the algorithm of \cite{neidinger2009multivariable}
to map into the Newton basis, then expand into the monomial basis.  It is possible to perform all computations required for polynomial interpolation using integers only; see \autoref{polynomial}.
To avoid a significant slowdown due to memory allocation inside GMP, the final version was written using manual memory allocation and the low level interface to GMP.

In addition to computing the perturbed signs of polynomial predicates, we use our scheme to compute exactly rounded perturbed constructions.  Given a rational function $f(x)/g(x)$
with $g(x) = 0$, we compute the perturbation series $g_1, g_2, \ldots$ until we find a nonzero $g_k$, compute the perturbed numerator $f_k$, then evaluate the perturbed result as
the ratio of the matching least infinitesimal nonzero term in $f_k$ and $g_k$.  In a correct algorithm this ratio will always be finite, in that $f_k$ will never contain a nonzero term
larger than $g_k$, but it is easy to detect this case and throw an exception as an aid to debugging.  Note that the ratio of matching least infinitesimal terms is exactly l'H\^opital's
rule for computing limits.  Finally, the ratio is rounded to the nearest integer.  We can similarly compute $\sqrt{f(x)/g(x)}$ by evaluating the limit of the ratio as a rational and
taking an exactly rounded square root.

We emphasize that these perturbed constructions are guaranteed to be within $L_0$ distance $1/2$ of the true answer, where the true answer is consistent with the rest of the algorithm and
obeys any geometric invariants that apply in the exact case.  For example, a constructed union of a convex polygon with itself will be within $L_0$ distance $1/2$ of the input, and in
particular will avoid all but extremely tiny foldovers that might result from performing constructions with floating point arithmetic when an algorithm completes.
Moreover, since the maximum error is known, they can be fed back into the same algorithm as tight interval bounds without fear of introducing inconsistencies.  Our
circular arc Boolean code makes use of this to perform more accurate interval-based filtering.  For example, when comparing $y$ coordinates of different intersections of circles,
we precompute the rounded intersections and avoid costly polynomial evaluation if the rounded coordinates differ.

Debugging and testing the symbolically perturbed algorithms we have implemented so far has been a quite pleasant experience.  Once the perturbation core itself is trusted,
bugs in the surrounding algorithm necessarily manifest on a set of positive measure, since any taken branching path through the code is described by algebraic inequalities which
give rise to open sets.  Thus, all bugs are likely to be found by running the algorithm on random input.  In contrast, an
algorithm which handles degeneracies specially or tailors the perturbation to the predicates involved must actually test each kind of degeneracy when debugging the algorithm.
Any speedup logic such as interval filtering can be easily checked by including a compile time flag to unconditionally evaluate both fast and slow paths.  This tests both the
correctness of the filter and the correctness of the predicate, which is important for complicated predicates.

Although our currently implemented algorithms are serial, our symbolic perturbation scheme can easily be used in parallel algorithms since each predicate evaluation is deterministic.
However, the dramatic slowdown between interval filtering and perturbed exact evaluation might interfere with load balancing at very high levels of parallelism, such as on a GPU.

In a correct geometric algorithm, no polynomial passed to symbolic perturbation will be identically zero; this would correspond to a fundamentally degenerate question such as
``Is the triangle $(x_7,x_7,x_7)$ counterclockwise?''.  However, it is convenient for debugging to detect these cases and produce useful output.  Therefore,
if both $f_1$ and $f_2$ are identically zero, our code pauses to run a randomized polynomial identity check \cite{schwartz1980fast} and throws an exception if a nonzero is not found.
The identity test evaluates the polynomial on 20 random points; this produces a false positive with probability under $10^{-171}$ (sufficient for the
lifetime of the code) and always reports failure for a truly zero polynomial.  The check has negligible effect on overall cost, since usually $f_1 \ne 0$.

For Delaunay triangulation, we use the partially randomized incremental construction of \cite{amenta2003incremental}.  Our implementation is $O(n \log n)$ for arbitrarily degenerate input,
and happily computes a random but valid Delaunay triangulation if all points are at the origin.  For Boolean operations, we find intersections using axis-aligned bounding box hierarchies
and find winding numbers for each contour by tracing rays along horizontal lines (horizontal lines are safe due to symbolic perturbation).  Our current Boolean
operation algorithms degrade to $O(n^2 \log n)$ for fully degenerate input since they compute an arrangement of curves as the first step; this slowdown is independent of the perturbation
technique used, and also occurs for badly formed nondegenerate input.  Compared to \cite{devillers2000algebraic}, which used degree 12 predicates for circular arc arrangements,
our implementation uses predicates of degree at most 8 via a combination of polynomial factoring and algorithmic changes (see \autoref{sec:predicates}).  Even degree 8 is problematic
for Yap's scheme due to the worst case exponential blowup in the number of terms.  Other work on circle arrangements in CGAL was done by \cite{wein2006circles}; this is orthogonal to our contribution.

\section{Results} \label{sec:results}

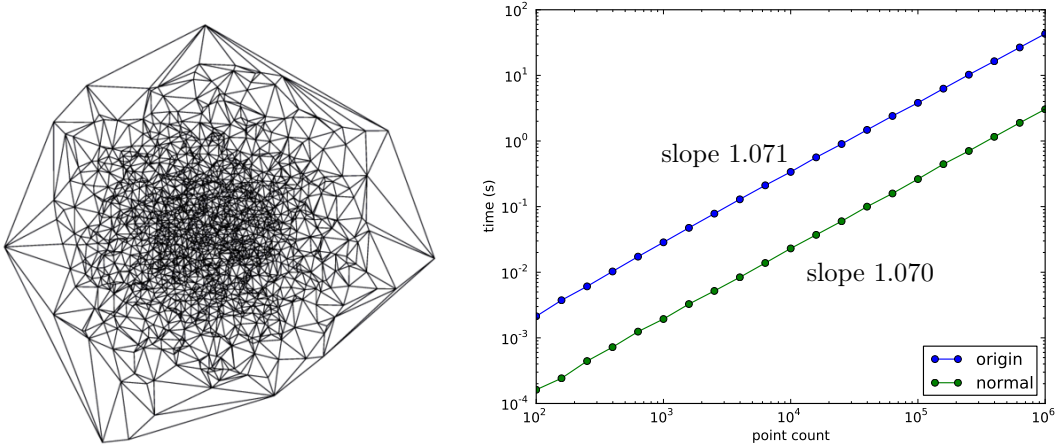
\begin{figure}
\centering
\includegraphics[height=.37\columnwidth]{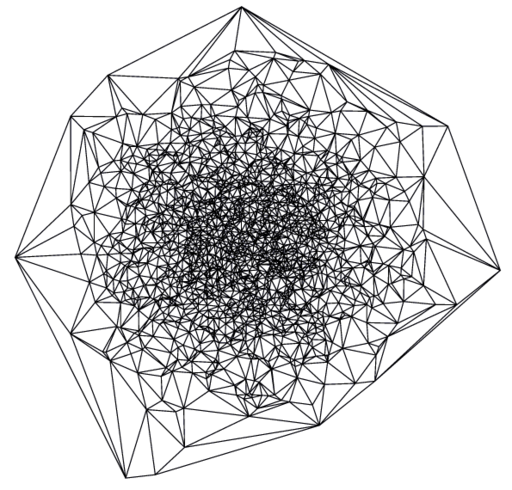}
\includegraphics[height=.42\columnwidth]{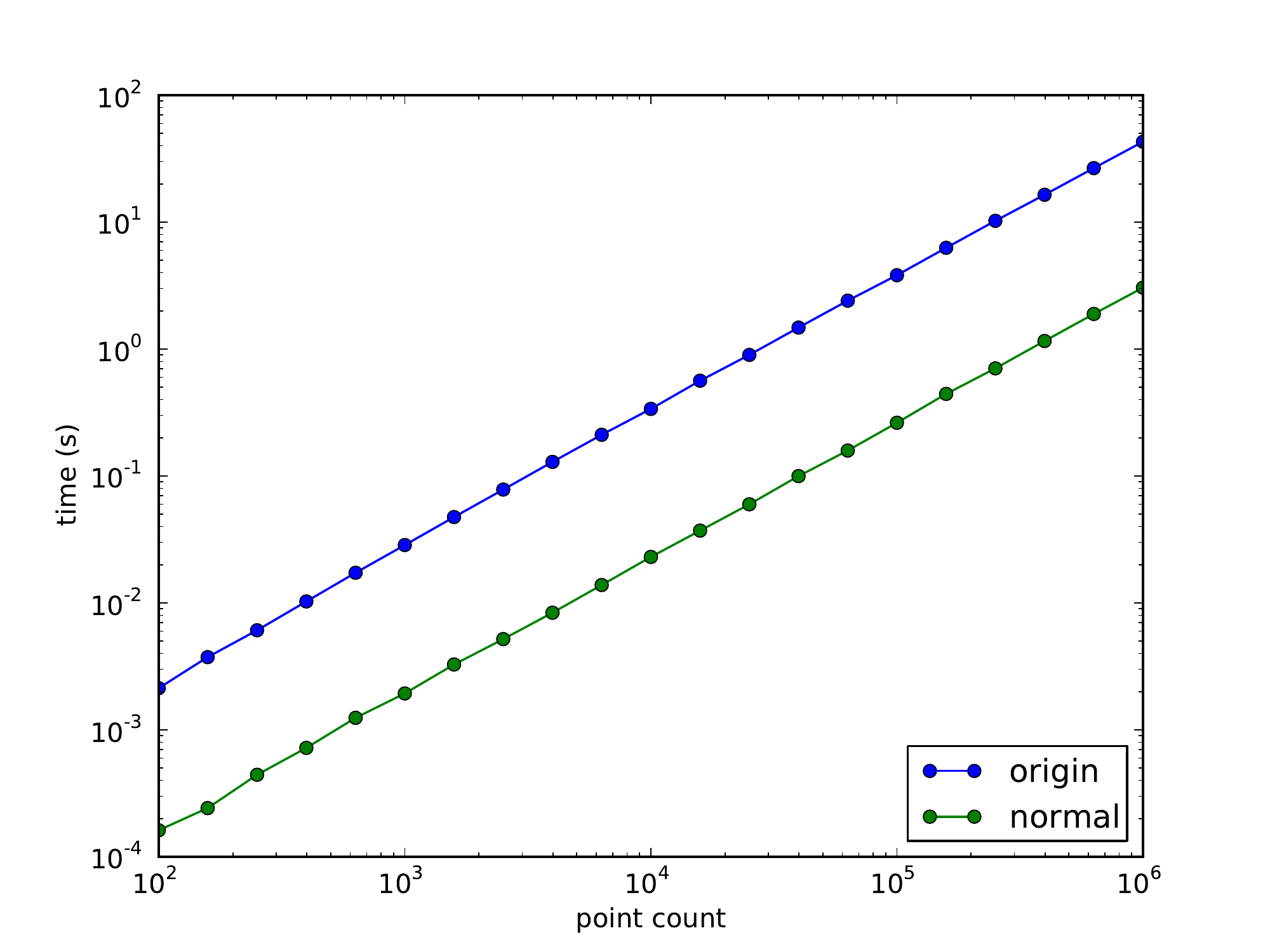}%
\begin{picture}(0,0)
\put(-170,110){{\small slope $1.071$}} 
\put(-115,65){{\small slope $1.070$}}  
\end{picture}
\cprotect\caption{Left: Delaunay triangulation of 2000 normally distributed points.  Right: computation time for Delaunay triangulation of (green, lower) $n$ normally distributed points
and (blue, upper) $n$ copies of the origin.  The fully degenerate case ranges from $13.1$ to $15.5$ times as slow as the random case due to falling back from interval
arithmetic filters to integer computation and symbolic perturbation.  To reproduce these figures, run \verb+examples delaunay --count 2000 --plot 1+ and
\verb+examples delaunay --count 1000000+.}
\label{fig:delaunay}
\end{figure}

Results for Delaunay triangulation are shown in \autoref{fig:delaunay}.  Since our algorithm is worst case $O(n \log n)$ independent of degeneracies, the slowdown ratio from random
input to fully degenerate input (all points at the origin) is constant: between $13$ and $15.5$ due to falling back from interval arithmetic filters to exact integer computation
and symbolic perturbation.  We note that our current Delaunay triangulation algorithm is not state of the
art, though this is orthogonal to our contributions: CGAL's routine is 4.3 times faster on $10^6$ normally distributed points ($0.704$ s vs. $3.05$ s).  It is also dramatically
faster for all points at the origin ($0.11$ s vs. $43$ s), though only because CGAL prunes duplicate points as a preprocess.  To reproduce our CGAL benchmarks, run
\verb+examples delaunay --count 1000000 --cgal 1+.

\begin{figure}
\centering
\includegraphics[height=.37\columnwidth]{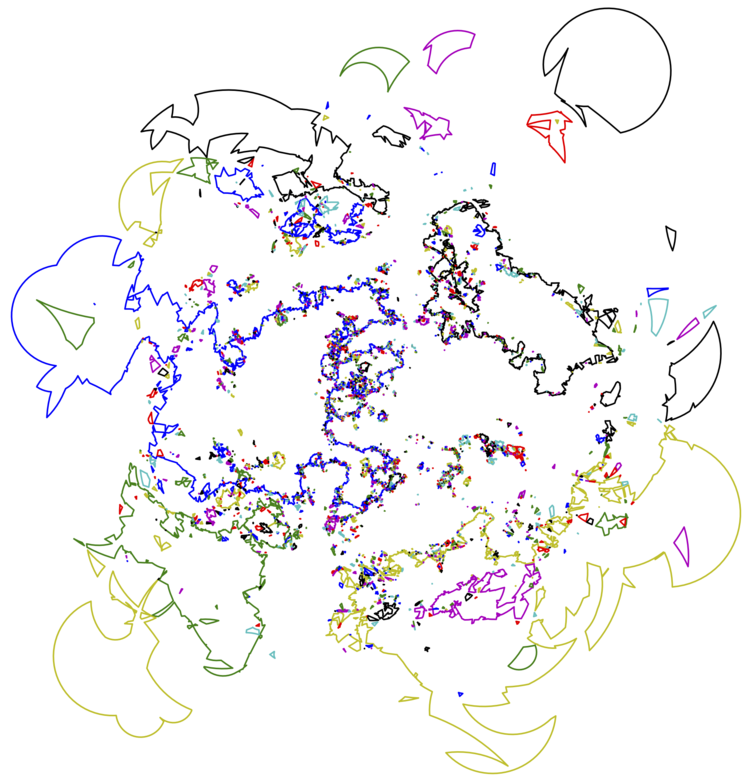}
\includegraphics[height=.37\columnwidth]{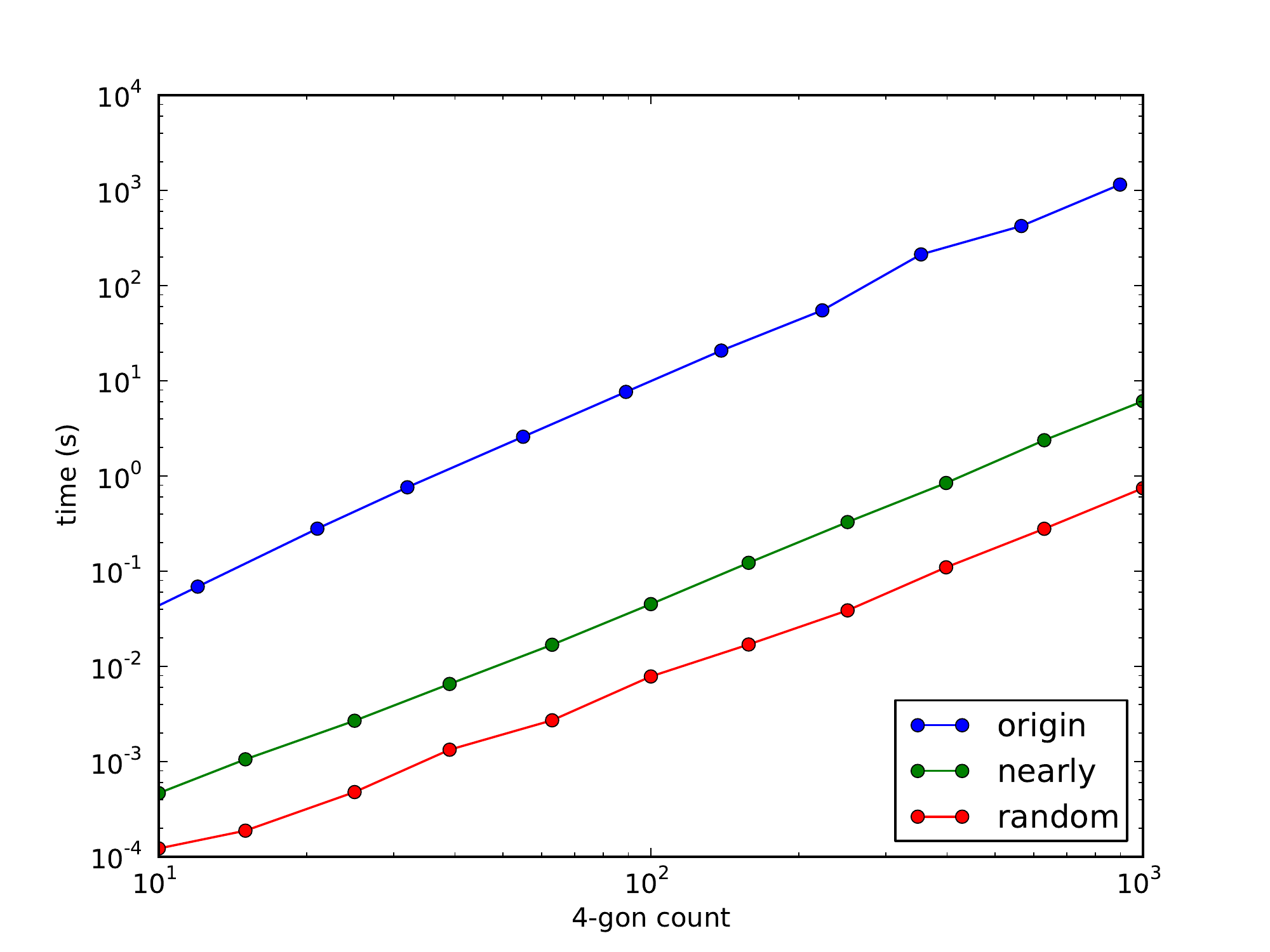}
\begin{picture}(0,0)
\put(-150,105){{\small slope $2.275$}} 
\put(-145,70){{\small slope $2.069$}}  
\put(-135,30){{\small slope $1.918$}}  
\end{picture}
\cprotect\caption{Left: Boolean union of 1000 randomly chosen circular arc 4-gons.  Right: computation time for union of different numbers of (red, lower) randomly distributed
4-gons, (green, middle) nearly but not exactly degenerate 4-gons, and (blue, top) exactly degenerate 4-gons.  The exactly degenerate case ranges from 65 to 252 times slower
than the nearly degenerate case, which is as expected since most of the cost is in degree 6 or 8 predicates ($6^3 = 216$, $8^3 = 512$).  Both random and nearly degenerate cases
use almost entirely interval arithmetic; the latter is slower since it is closer to the quadratic worst case.  To reproduce these figures, run
\verb+examples circles --plot 1 --count 1000+ and \verb+examples --mode circles --count 1000 --min-count 10+.}
\label{fig:circles}
\end{figure}
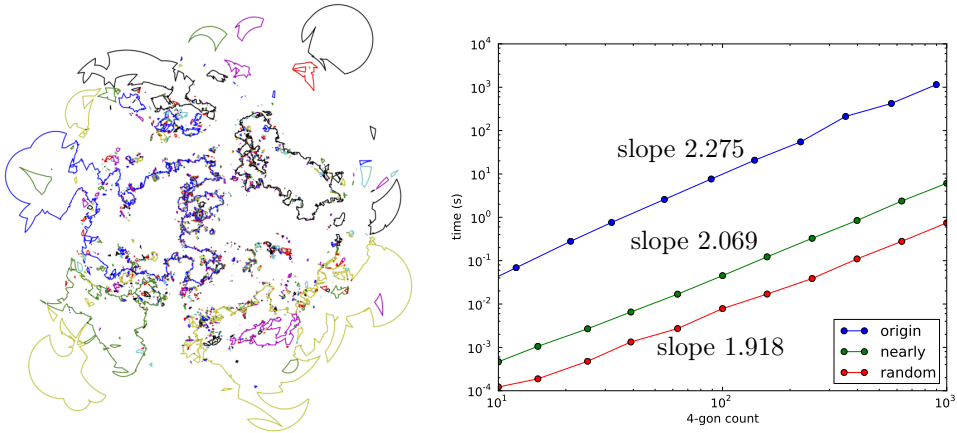

Results for circular arc Booleans are shown in \autoref{fig:circles}.  Log-log slopes near 2 are expected because of the $O(n^2)$ complexity of general arrangements of circles.
The slowdown for the exactly vs.\ nearly degenerate case is much greater than for Delaunay triangulation because of the higher degree and increased complexity of the predicates.
Further optimizations to the degenerate case are possible, in particular inlining GMP calls for small arguments and caching certain repeated predicate evaluations, but these are
of questionable importance in practice since a tiny amount of finite jittering removes the vast majority of degeneracies.

\section{Conclusion}

We have presented a deterministic pseudorandom symbolic perturbation scheme which combines the advantages of several existing techniques.  Given a polynomial $f(x)$, we evaluate
the sign of $f(x + \epsilon_1 y_1 + \epsilon_2 y_2 + \cdots)$ where $y_k$ are deterministic pseudorandom and $\epsilon_k$ are infinitesimals in decreasing order of size.
Typically only the first infinitesimal in this series need be considered, so our method is as fast as the linear symbolic perturbation schemes, but works for arbitrary
polynomials and appears deterministic to the caller.

\bibliography{references}
\bibliographystyle{acm}
\appendix

\section{Polynomial interpolation} \label{polynomial}

We found several useful papers discussing different aspects of univariate and multivariate polynomial interpolation, and collect these results for convenience.  The algorithms
discussed here perform $O(N^2)$ linear operations to convert $N$ samples to $N$ coefficients.  Adds and multiply-by-constants for degree $d$ integers require time $O(d)$, so
the total complexity is $O(d N^2)$.  Asymptotically faster algorithms using spectral methods exist, but we do not consider them here.

In order to recover the coefficients of $f_k(\epsilon_1, \ldots, \epsilon_k)$ we must perform multivariate interpolation given the values of $f_k$ at our chosen set of tuples.
In the univariate case, this amounts to the classical divided difference algorithm.  As discussed in \cite{oruc2000explicit} and \cite{olver2006multivariate}, the divided
difference algorithm can be beautifully expressed as the following factorization of the Vandermonde matrix into bidiagonal matrices, shown here for the degree 3 case:
\begin{equation} \label{vandermonde}
\begin{aligned}
\left(\begin{matrix}1 & x_{0} & x_{0}^{2} & x_{0}^{3}\\1 & x_{1} & x_{1}^{2} & x_{1}^{3}\\1 & x_{2} & x_{2}^{2} & x_{2}^{3}\\1 & x_{3} & x_{3}^{2} & x_{3}^{3}\end{matrix}\right)
=& \left(\begin{matrix}1&0&0&0\\\frac{1}{x_{0}-x_{1}}&\frac{1}{x_{1}-x_{0}}&0&0\\0&\frac{1}{x_{1}-x_{2}}&\frac{1}{x_{2}-x_{1}}&0\\0&0&\frac{1}{x_{2}-x_{3}}&\frac{1}{x_{3}-x_{2}}\end{matrix}\right)^{-1} \\
 & \left(\begin{matrix}1&0&0&0\\0&1&0&0\\0&\frac{1}{x_{0}-x_{2}}&\frac{1}{x_{2}-x_{0}}&0\\0&0&\frac{1}{x_{1}-x_{3}}&\frac{1}{x_{3}-x_{1}}\end{matrix}\right)^{-1} \\
 & \left(\begin{matrix}1&0&0&0\\0&1&0&0\\0&0&1&0\\0&0&\frac{1}{x_{0}-x_{3}}&\frac{1}{x_{3}-x_{0}}\end{matrix}\right)^{-1} \\
 & \left(\begin{matrix}1&x_{0}&0&0\\0&1&x_{1}&0\\0&0&1&x_{2}\\0&0&0&1\end{matrix}\right) \\
 & \left(\begin{matrix}1&0&0&0\\0&1&x_{0}&0\\0&0&1&x_{1}\\0&0&0&1\end{matrix}\right) \\
 & \left(\begin{matrix}1&0&0&0\\0&1&0&0\\0&0&1&x_{0}\\0&0&0&1\end{matrix}\right) 
\end{aligned}
\end{equation}
This factorization was given in \cite{oruc2000explicit}, though in a somewhat less elegant form due to placing ones along the diagonal of $L$ instead of $U$ in the $LU$ factorization.
The clean $LU$ factorization was given in \cite{olver2006multivariate}, though without the further bidiagonal factorization.

The first half of this factorization is the classical divided difference algorithm to convert values $f(x_0), \ldots, f(x_k)$ into the coefficients of $f$ in the Newton
basis $x(x-1)\cdots(x-n+1)$.  The second half expands from the Newton basis down to monomials.  In our case, we have $x_k = k$, so all of the ratios in each bidiagonal matrix have the same denominator.
In particular, we can clear fractions by multiplying the inverse by $d!$ where $d$ is the degree of $f$, after which all computations can be performed in integers.  Alternatively,
we can use the fact that while the inverse of the Vandermonde matrix is not integral, both our polynomial values and the coefficients of the polynomials in both Newton and monomial basis
are integers.  It turns out that in this case all intermediate results in the divided difference algorithm are integers as well.  To show this, we must prove that the $k$th forward
difference $\Delta^k f(x)$ of an integer polynomial is divisible by $k!$.  We use the following argument due to Qiaochu Yuan\footnote{\url{http://math.stackexchange.com/questions/413600}}.
Since the transformation to and from the monomial basis to Newton basis (the second half of (\ref{vandermonde})) is integral, it suffices to check $k! \mid \Delta^k f(x)$ for an
element of the Newton basis
\begin{linenomath*}
$$f(x) = x(x-1)\cdots(x-n+1) = n!\binom{x}{n}.$$
\end{linenomath*}
Since $\Delta \binom{x}{n} = \binom{x}{n-1}$ we have
\begin{linenomath*}
\begin{align*}
\Delta^k x(x-1)\cdots(x-(n-1))
  &= n! \binom{x}{n-k} \\
  &= \frac{n!}{(n-k)!} x(x-1)\cdots(x-(n-k-1)) \\
  &= k! \binom{n}{n-k} x(x-1)\cdots(x-(n-k-1))
\end{align*}
\end{linenomath*}
For the multivariate case, Neidinger \cite{neidinger2009multivariable} provides an elegant generalization of the univariate divided difference algorithm when the polynomial is
evaluated on an ``easy corner'' of points, which includes the $0 \le \epsilon_i$, $\epsilon_1 + \cdots + \epsilon_k \le d$ set that we use.  All intermediate results
in their algorithm are multivariate divided differences and are therefore integral by the above argument.  They discuss only interpolation into the multivariate Newton basis
consisting of polynomials such as
\begin{linenomath*}
$$\prod_i x_i(x_i-1)\cdots(x_i-(n_i-1))$$
\end{linenomath*}
which corresponds to the first half of \autoref{vandermonde}.  The multivariate generalization of the second half of \autoref{vandermonde} is easy, since the multivariate Newton to
monomial basis transformation matrix factors into commuting matrices each expanding one variable, and these matrices are block diagonal with respect to the other variables.

\section{Degree 8 circular arc predicates} \label{sec:predicates}

The critical predicate required for circular arc arrangements, determining whether one intersection of two arcs is above another intersection,
can be reduced to degree 12 using resultant techniques \cite{devillers2000algebraic}.  This holds for the general case of two unrelated intersections between pairs of circles
$C_0,C_1$ and $C_2,C_3$.  However, to compute a circular arc arrangement it suffices to consider the case where $C_0 = C_2$; that is, comparing the $y$ coordinates of the
intersections of one circle with two others.  In this case, the polynomials can be factored into terms of degree $\le 8$.  One significant algorithmic change is required,
since we can no longer fire a horizontal or vertical ray from the intersection of $C_0,C_1$ and detect intersections against unrelated circle arcs.  Instead, we must fire
rays along exactly known (degree 1) $y$ coordinates, which is sufficient to determine the winding number of a given circular arc polygon (or connected component of an arrangement)
as long as the bounding box touches at least one ray.  For most applications, polygons smaller than this may be safely discarded.

We derived the degree 8 version of the predicate by starting with an inequality involving square roots, then iteratively checking polynomial signs and squaring to eliminate
square roots until a fully polynomial inequality is reached.  All polynomials to be tested were then factored in Mathematica down to their minimal degree, then manually simplified
down to the more compact expressions shown below (Mathematica's \verb+FullSimplify+ was insufficient for this purpose), using Mathematica to check each stage of the simplification.
The resultant techniques used in \cite{devillers2000algebraic} would have also found the degree 8 solution had they been applied to the three circle special case.  It should be possible
to automate the entire process from algebraic inequality to optimized minimum degree polynomial expressions, but we have not yet done so.

The derivations below make several simplifications, for example assuming that squaring does not reverse the direction of inequalities.  For full details, refer to
\url{https://github.com/otherlab/core/blob/b186ab68303/exact/circle_predicates.cpp#L289} or \verb+circles.nb+ in \url{https://github.com/otherlab/perturb}.

\subsection{The intersection of two circles}

Let circle $C_i$ have center $c_i$ and radius $r_i$, and define $c_{ij} = c_j - c_i$.  Assuming $C_0$ and $C_1$ intersect, parameterize one of their intersections by
\begin{linenomath*}
\begin{align*}
p_{01} &= c_0 + \alpha c_{01} + \beta c_{01}^\perp.
\end{align*}
\end{linenomath*}
where $v^\perp$ is $v$ rotated left by $90^\circ$.  We have
\begin{linenomath*}
\begin{align*}
(p_{01} - c_i)^2 &= r_i^2 \\
p_{01}^2 - 2p_{01} \cdot c_i + c_i^2 &= r_i^2.
\end{align*}
\end{linenomath*}
Subtracting the two circle equations gives
\begin{linenomath*}
\begin{align*}
-2p_{01} \cdot c_{01} + c_1^2 - c_0^2 &= r_1^2 - r_0^2 \\
-2c_0 \cdot c_{01} -2\alpha c_{01}^2 + (c_0 + c_1) \cdot c_{01} &= r_1^2 - r_0^2 \\
(1-2\alpha) c_{01}^2 &= r_1^2 - r_0^2 \\
1 - 2 \alpha &= \frac{r_1^2 - r_0^2}{c_{01}^2} \\
\halpha = 2 c_{01}^2 \alpha &= c_{01}^2 + r_0^2 - r_1^2
\end{align*}
\end{linenomath*}
Substituting into $C_0$'s equation gives
\begin{linenomath*}
\begin{align*}
(p_{01} - c_0)^2 &= r_0^2 \\
\left(\alpha c_{01} + \beta c_{01}^\perp \right)^2 &= r_0^2 \\
\alpha^2 c_{01}^2 + \beta^2 c_{01}^2 &= r_0^2 \\
\beta^2 &= \frac{r_0^2}{c_{01}^2} - \alpha^2 \\
\hbeta^2 = \left(2 c_{01}^2 \beta\right)^2 &= 4 r_0^2 c_{01}^2 - \halpha^2.
\end{align*}
\end{linenomath*}
To summarize, the intersection between circles $C_0$ and $C_1$ is described by
\begin{linenomath*}
\begin{align*}
p_{01} &= c_0 + \alpha c_{01} + \beta c_{01}^\perp \\
\halpha = 2 \alpha c_{01}^2 &= c_{01}^2 - r_1^2 + r_0^2 \\
\hbeta^2 = (2c_{01}^2 \beta)^2 &= 4 r_0^2 c_{01}^2 - \halpha^2
\end{align*}
\end{linenomath*}
where we choose the positive or negative square root for $\beta$ depending on which intersection is desired.

\subsection{Is one circle intersection above another?}

Given three circles $C_0,C_1,C_2$, is $p_{01}$ below $p_{02}$?  This predicate has the form
\begin{linenomath*}
\begin{align*}
p_{01y} &< p_{02y} \\
c_{0y} + \alpha_{01} c_{01y} + \beta_{01} c_{01x} &< c_{0y} + \alpha_{02} c_{02y} + \beta_{02} c_{02x} \\
0 &< \alpha_{02} c_{02y} - \alpha_{01} c_{01y} - \beta_{01} c_{01x} + \beta_{02} c_{02x} \\
0 &< \halpha_{02} c_{02y} c_{01}^2 - \halpha_{01} c_{01y} c_{02}^2 - \hbeta_{01} c_{01x} c_{02}^2+ \hbeta_{02} c_{02x} c_{01}^2 \\
0 &< A + B_1 \sqrt{C_1} + B_2 \sqrt{C_2}
\end{align*}
\end{linenomath*}
where $A,B_1,B_2,C_1,C_2$ are polynomials and $C_1, C_2 > 0$ since the two intersections are assumed to exist.  To reduce this equality to
purely polynomial equalities, we first compute the signs of $A, B_1, B_2$.  If these all match, we are done.  Otherwise we move the square root
terms that differ from $A$ in sign to the RHS and square.  Assuming $A > 0$, this gives either
\begin{linenomath*}
\begin{align}
A + B_1 \sqrt{C_1} &> -B_2 \sqrt{C_2} \nonumber \\
A^2 + B_1^2 C_1 + 2 A B_1 \sqrt{C_1} &> B_2^2 C_2 \nonumber \\
A^2 + B_1^2 C_1 - B_2^2 C_2 &> -2 A B_1 \sqrt{C_1} \label{one-b}
\end{align}
\end{linenomath*}
or
\begin{linenomath*}
\begin{align}
A &> -B_1 \sqrt{C_1} - B_2 \sqrt{C_2} \nonumber \\
A^2 &> B_1^2 C_1 + B_2^2 C_2 + 2 B_1 B_2 \sqrt{C_1 C_2} \nonumber \\
A^2 - B_1^2 C_1 - B_2^2 C_2 &> 2 B_1 B_2 \sqrt{C_1 C_2} \label{two-b}
\end{align}
\end{linenomath*}
The signs of the RHS's of (\ref{one-b}) and (\ref{two-b}) are known.  The polynomial LHS's are degree 10, but factor as
\begin{linenomath*}
\begin{align*}
\begin{array}{@{}r} A^2 + B_1^2 C_1 - B_2^2 C_2 \phantom{\bigg(} \hspace{-.5em}= \\ \phantom{\bigg(} \end{array}&%
\begin{array}{@{}r@{}l@{}} c_{02}^2 \bigg(& c_{01}^2 \left(\halpha_{02} \left(\halpha_{02} c_{01}^2 - 2 \halpha_{01} c_{01y} c_{02y}\right)
  + 4 r_0^2 (c_{01x}^2 c_{02y}^2 - c_{01y}^2 c_{02x}^2)\right) \\
  &- \halpha_{01}^2 \left(c_{01x}^2 - c_{01y}^2\right) c_{02}^2 \bigg) \end{array} \\
\begin{array}{@{}r} A^2 - B_1^2 C_1 - B_2^2 C_2 \phantom{\bigg(} \hspace{-.5em}= \\ \phantom{\bigg(} \end{array}&%
\begin{array}{@{}r@{}l@{}} c_{01}^2 c_{02}^2 \bigg(& c_{02}^2 \halpha_{01}^2 + c_{01}^2 \halpha_{02}^2 - 2 c_{01y} c_{02y} \halpha_{01} \halpha_{02} \\
  &- 4 r_0^2 (c_{01y}^2 c_{02x}^2 + c_{01x}^2 c_{02y}^2 + 2c_{01x}^2 c_{02x}^2) \bigg) \end{array}
\end{align*}
\end{linenomath*}
and therefore reduce to degree 8 and 6, respectively.  If the LHS and RHS of (\ref{one-b}) or (\ref{two-b}) have the same sign, we square once
more to eliminate the final square root.  Assuming positive LHS, squaring (\ref{one-b}) gives
\begin{linenomath*}
\begin{align*}
(A^2 + B_1^2 C_1 - B_2^2 C_2)^2 &> 4 A^2 B_1^2 C_1 \\
A^4 - 2A^2 B_1^2 C_1 + B_1^4 C_1^2 - 2 A^2 B_2^2 C_2 - 2 B_1^2 B_2^2 C_1 C_2 + B_2^4 C_2^2 &> 0 \\
E &> 0 
\end{align*}
\end{linenomath*}
and squaring (\ref{two-b}) gives
\begin{linenomath*}
\begin{align*}
(A^2 - B_1^2 C_1 - B_2^2 C_2)^2 &> 4 B_1^2 B_2^2 C_1 C_2 \\
A^4 - 2A^2 B_1^2 C_1 + B_1^4 C_1^2 - 2 A^2 B_2^2 C_2 - 2 B_1^2 B_2^2 C_1 C_2 + B_2^4 C_2^2 &> 0 \\
E &> 0.
\end{align*}
\end{linenomath*}
That is, the two inequalities square into the same degree 20 polynomial $E$, which factors into degree $\le 6$ terms as
\begin{linenomath*}
\begin{align*}
E &= c_{01}^4 c_{02}^4 E_+ E_- \\
E_\pm &= c_{02}^2 \halpha_{01}^2 + c_{01}^2 \halpha_{02}^2 - 2 \halpha_{01} \halpha_{02} (c_{01y} c_{02y} \pm c_{01x} c_{02x}) - 4 r_0^2 (c_{01x} c_{02y} \mp c_{01y} c_{02x})^2
\end{align*}
\end{linenomath*}
If intersections between four circles are compared, the analog to $E$ is still divisible by $c_{01}^4 c_{02}^4$, but the remaining degree $12$ polynomial
is irreducible as expected from \cite{devillers2000algebraic}.

As might be expected, performing these calculations only semiautomatically resulted in a large number of typos and copying errors.  The fact that the final
result is automatically checked against interval filters in the code was critical to making the debugging process practical.

\end{document}